\documentclass[prd,floatfix,preprintnumbers,letterpaper,preprint]{revtex4}
\usepackage{graphicx}
\usepackage{amsmath,amssymb}
\usepackage{epsfig}
\usepackage{latexsym}
\newcommand {\ga} {\ {\raise-.5ex\hbox{$\buildrel>\over\sim$}}\ }
\newcommand {\la} {\ {\raise-.5ex\hbox{$\buildrel<\over\sim$}}\ }

\begin{document}

\def\be{\begin{equation}}
\def\ee{\end{equation}}

\title{Late Time Decay of the False Vacuum, Measurement, and Quantum Cosmology }
\author{Lawrence M. Krauss$^{1}$, James Dent$^{2}$ and Glenn D. Starkman$^3$
\\krauss@asu.edu, james.b.dent@vanderbilt.edu, gds6@cwru.edu}
\affiliation{$^1$ School of Earth and Space Exploration and Physics Department, Arizona State University, Tempe, AZ 85287}
\affiliation{$^2$Department of Physics \& Astronomy, Vanderbilt University,
Nashville, TN~~37235}
\affiliation{$^3$CERCA, Department of Physics, Case Western Reserve University,
Cleveland, OH~~44106}

\date{\today}

\begin{abstract}
The recent suggestion that late time quantum dynamics may be important for resolving cosmological issues associated with our observed universe requires a consideration of several subtle issues associated with quantum cosmology, as we describe here.  The resolution of these issues will be important if we are to be able to properly ascribe probability measures associated with eternal inflation, and a string landscape.
\end{abstract}

\maketitle

Applying quantum mechanical arguments to the universe as a whole is notoriously tricky.  The difficulties do not necessarily arise at a mathematical level, but rather in the interpretation of the mathematics.   Indeed, while some computations associated with the wavefunction of the universe may be well defined, interpreting the results  is much less clear, given the dual facts that (a) we are not external observers, and (b) we only probe a single universe from the enormous space of classical universes embodied in the quantum wavefunction. 

We have already come face to face with the inadequacy of having only one universe with which to probe fundamental physics.  Measurements of the low multipoles of the Cosmic Microwave Background, for example, are subject to "cosmic variance" limits, implying that we can probe only a limited number of independent degrees of freedom chosen from a gaussian random sample (if indeed primordial density fluctuations arise from quantum mechanical vacuum fluctuations in the early universe), and therefore no matter how precisely we measure the CMB, we have a limited ability to observationally characterize the distribution from which the data may be drawn.

Another area where quantum mechanical arguments, and the ambiguity of measurements associated with our universe becomes significant involves vacuum decay, in particular the possibility, as suggested by chaotic inflation that our observable universe  is a very small part of a much larger ensemble of universes, or multiverse, in which metastable vacuum decays occurred and may still be occurring. 

In this case it becomes likely that our entire observable universe arose out of a quantum mechanical fluctuation associated with the decay of a meta-stable false vacuum configuration in the early universe.   Moreover, if the manifold of vacuum states is sufficiently complex the parameters of our universe are probably stochastically determined by which specific false vacuum state decayed into our state and by the subsequent state in which our universe resides. 

If one could enumerate all possible vacuum candidate configurations, and also calculate their decay rates into all accessible configurations, one might hope to assign a probability measure to the possible outcomes.  Even then, however, one would need to take into account another crucial factor, namely the volume of space-time occupied by each accessible configuration, which in turn requires incorporating the gravitational effects associated with false vacua, especially superluminal expansion. 

Recently \cite{kraussdent} two of us have argued that the late-time properties associated with the  quantum mechanical decay of metastable states can add a new wrinkle to efforts to infer the properties of the observed universe on the basis of probability estimates in a multiverse.  For any quantum mechanical system with a fixed lowest energy state, arguments based on analyticity imply that quantum mechanical decay cannot remain exponential indefinitely.  Power law behavior must take over at late times, and dimensional arguments suggest that a crossover time will depend logarithmically on ratio of the only two dimensional parameters in the problem -- the energy gap between states, and the decay rate.

If one adapts these arguments to the case of false vacuum decay, one can derive an estimate \cite{kraussdent} for the crossover time, $T$, in terms of the ratio between the energy density gap, $E_o$, between false and true vacua, and the decay rate per unit volume, $\Gamma$ as:

\begin{eqnarray}
\label{largisher time}
T \sim \frac{O(5)}{\Gamma}ln\bigg(\frac{E_o}{\Gamma}\bigg) 
\end{eqnarray}   

This implies that the survival probability (per unit volume) at this time would be given by:
 \begin{eqnarray}
\label{largish time}
{N \over N_0} \approx \bigg(\frac{\Gamma}{E_o}\bigg)^{5}
\end{eqnarray}

The implications of this result for the calculation of probability measures in an eternally inflating landscape could be significant.   False vacua typically inflate exponentially. If the transitions out of the false vacuum percolate in a time much shorter than $T$, then the probability of finding a given region of space in the false vacuum at intermediate times will not change.  However, if the transitions do not percolate, or for times longer than $T$, when the decay rate per unit volume in those regions in which false vacua have persisted is no longer exponential, the calculation of relative volumes of space residing in metastable false vacuum configurations will change considerably.

For example, because the crossover time depends upon $E_o$, those vacua with the smallest energy gaps  for a given decay rate will have an exponentially higher likelihood of survival as a function of time than those with larger gaps.  This can serve to counterbalance either completely or partially the increased total volume of space normally assigned in the probability measure to false vacua with larger gaps due to the fact that they expand with a larger exponential factor.  Moreover, because each such volume has an exponentially larger decay, if one convolves the probability measure based on total volume with the likelihood of living long enough for structure to form, the net probability may become dominated by false vacua with the lowest allowed energy gap.  Could such an argument help provide an anthropic explanation for the observed smallness of the observed vacuum energy in our universe?

Moreover, when one considers life-forms at late times, power law late time vacuum decay will also exacerbate the so-called Boltzmann Brain problem \cite{kraussstarkdent}.   For example, even if one attempts to make inflating vacua unstable in a string landscape, one may find that the metastable configurations still dominate in spatial averaging for all times, so that thermal fluctuations in such configurations may still inevitably come to dominate over other life-forms in these spaces. 

Implicit in these conclusions however, is the assumption that large inflating universes can persist with non-zero probabilities, for timescales that can be long compared to their characteristic decay times.  It is this question that we wish to elaborate upon here.

First, working against the power-law late-time persistence of metastable quantum configurations can be the so-called Quantum Zeno effect.  Stated simply, this effect can result from the fact that if quantum systems are measured on a timescale shorter than their characteristic decay time, their 'decay clocks' are reset as the wavefunction of the system continues to remain fixed near a pure metastable state.  In the limit of continuous successive and sufficiently frequent observations, the system could then be maintained in a metastable state that does not decay.   More generally, if they are repeatedly measured on longer timescales, so the wavefunction is in the late-time decay phase, the nature of late time decay can itself change as well \cite{sudarshan}.  The form of the variation away from either the canonical exponential decay law for intermediate times, or the long time power law can be determined simply as follows.

Consider the survival probability $Q(t)$, whose standard intermediate time exponential form is $\exp{-\Gamma t}$ .  For very short times, it will depart from this form and can be
written

\begin{equation}
Q(t) \approx 1- \kappa t^{\beta}
\end{equation}

If $\beta > 1$ then if one makes $n$ repeated measurements over time intervals $t/n$ then $Q(t)^n$ approaches unity as n becomes arbitrarily large for finite $t$.   This reflects the so-called quantum Zeno effect. 

Consider on the other hand what repeated measurements of the system at longer time intervals might do.  Because each such measurement collapses the wavepacket, one must restart the decay clock, so that in general if measurements are made at regular intervals $t$, then this will imply that 
$Q(nt) = Q(t)^n$.    If $n$ is large this implies reversion to an exponential decay law once again, independent of whether $Q(t)$ itself has begun to depart from an exponential form.

As a result, we will only expect a system to undergo late-time power law decay if the system is ``closed", that is, not subject to external measurement.  On the other hand, even an unstable system can become arbitrarily stable if measurements are performed frequently enough. 

Clearly the implications of these effects on the impact of false vacuum decay on the nature of an eternally inflating multiverse could be significant.  If the false vacuum configuration is continually 'measured', could one effectively preserve a false vacuum configuration indefinitely?   And if measurements are made at intermediate times, can one ensure that one never effectively reaches the power-law late-time decay regime?

From one perspective, it would seem that, by definition, causally disconnected false vacuum regions are closed systems which therefore must evolve globally via quantum mechanics, and thus are destined, at late times, to experience power law decay properties.  However, this issue becomes murkier if one considers the possibilities that these regions evolve to become large, so that they can contain classical internal observers, and so that macroscopic objects like galaxies may form.

Arguably, no internal observation of a false vacuum region will collapse its global wavefunction.  Observations made by internal observers will simply entangle these observers with the regions within the false vacuum that they have observed.  The entire closed system will continue to evolve quantum mechanically, with entangled sub-components.  

The problem arises when we consider the exact physical nature of false vacuum decay.   If this occurs, for example, via bubble nucleation, the initial size of the bubble will be governed by microscopic physical parameters, and will essentially be a local process.  We can ask, for example, whether propagation of a dilute gas of photons through such a 'small' region (dilute enough to have a negligible effect on the overall expansion of the region) may comprise an observation sufficient to collapse the wavefunction of this region to its initial pure false vacuum state.  Such photons, for example, in an inflating background will experience an exponential redshift that can be measured.  While from a global perspective this may be considered to merely entangle the observer with the quantum mechanical configuration of the measured region, nevertheless if such a region is continually forced back into its initial state, then might we not expect its subsequent decay probabilities to be altered as described above depending upon the average time between measurements?

It is important here that the classical observation can actually distinguish a pure false vacuum state from its subsequently evolved configuration, which, if it is metastable, is actually in a superposition of inflating and non-inflating states.   Otherwise probing the system in this way will have no effect on its subsequent evolution.  Measuring the redshift of photons traveling through space however on the surface seems to be just such an observation.  

Clearly this question is of vital interest for determining the overall dynamics of an eternally inflating multiverse, and hence for estimating probability measures therein.   As described at the beginning of this essay, the mathematics governing the different dynamical possibilities is straightforward, but the interpretation of this mathematics in terms of the quantum wavefunction of our universe is trickier. 
If global evolution of causally disconnected universes remains quantum mechanical, then late-time power-law decay will be the dominant process affecting the overall properties of space at late times.  If, on the other hand, the behavior of local sub-systems within these universes, which are not themselves closed systems, is important for determining the dynamics of false vacuum decay, then late time decay may no longer be important.  However even in this case, it is not clear what the result will be.  Either exponential decay will be preserved, or, perhaps, frequently interacting subsystems can remain pseudo-stable indefinitely via the Zeno effect. 

Our purpose here is to raise the importance of resolving this key issue, which can have a dramatic impact on the possible implications, anthropic and otherwise, for determining the evolution of an eternally inflating multiverse.

\end{document}